\documentclass[%
aip,
rsi,%
amsmath,amssymb,
reprint,%
]{revtex4-1}

\usepackage{graphicx}% Include figure files
\usepackage{dcolumn}% Align table columns on decimal point
\usepackage{bm}% bold math
\usepackage{algorithm}%
\usepackage{algorithmicx}%
\usepackage{algpseudocode}%
\usepackage{float}
\usepackage[colorlinks=true, linkcolor=blue, citecolor=blue, urlcolor=blue]{hyperref}

\begin{document}
	
	\title{A spiking photonic neural network of 40.000 neurons, trained with rank-order coding for leveraging sparsity}
	
\author{Ria Talukder}
\address{FEMTO-ST Institute/Optics Department, CNRS - University Franche-Comté, 15B avenue des Montboucons, Besançon Cedex, 25030, France}

\author{Anas Skalli}
\address{FEMTO-ST Institute/Optics Department, CNRS - University Franche-Comté, 15B avenue des Montboucons, Besançon Cedex, 25030, France}

\author{Xavier Porte}
\address{Institute of Photonics, Department of Physics, University of Strathclyde, 99 George str., Glasgow G1 1RD, UK}

\author{Simon Thorpe}
\address{Centre de Recherche Cerveau et Cognition CERCO UMR5549, CNRS—Université Toulouse III, Toulouse, France}

\author{Daniel Brunner}
\address{FEMTO-ST Institute/Optics Department, CNRS - University Franche-Comté, 15B avenue des Montboucons, Besançon Cedex, 25030, France}
\email{daniel.brunner@femto-st.fr}

	\date{\today}% It is always \today, today,
	
	\begin{abstract}
		
Spiking neural networks are neuromorphic systems that emulate certain aspects of biological neurons, offering potential advantages in energy efficiency and speed by for example leveraging  sparsity.
 While CMOS-based electronic SNN hardware has shown promise, scalability and parallelism challenges remain. Photonics provides a promising platform for SNNs due to the speed of excitable photonic devices standing in as neurons and the parallelism and low-latency of optical signal conduction.
 Here, we present a photonic SNN comprising 40,000 neurons using off-the-shelf components, including a spatial light modulator and a CMOS camera, enabling scalable and cost-effective implementations for photonic SNN proof of concept studies.
 The system is governed by a modified Ikeda map, were adding additional inhibitory feedback forcing introduces excitability akin to biological dynamics.
 Using latency encoding and sparsity, the network achieves 83.5\% accuracy on MNIST using 22\% of neurons, and 77.5\% with 8.5\% neuron utilization.
 Training is performed via liquid state machine concepts combined with the hardware-compatible SPSA algorithm, marking its first use in photonic neural networks.
 This demonstration integrates photonic nonlinearity, excitability, and sparse computation, paving the way for efficient large-scale photonic neuromorphic systems.
		
	\end{abstract}
	
	\maketitle
	
\section{Introduction}

Artificial neural networks (ANNs) have revolutionized the field of machine learning, showing remarkable success in various computational tasks, from image recognition \cite{lecun2015} to language processing \cite{vaswani2017}.
They are loosely inspired by the structure and function of the brain's neurons, with each neuron modeled as a usually highly simplified mathematical unit, the perceptron \cite{rosenblatt1958}.
This abstraction has been incredibly effective, mostly when such models are applied to computational tasks, however slightly less so when applied to computational neuroscience where ANNs are to aid in understanding biological processes.
Conventional ANNs deviate significantly from the mechanisms observed in biological neural networks, and as such also fail in efficiently exploiting hardware in a similarly, more neuromorphic way \cite{indiveri2011}.
Spiking neural networks (SNNs) potentially represent a closer approximation to biological systems \cite{MAASS1997}.
Neurons in SNNs only spike when an input exceeds a certain threshold, mirroring the way neurons in the brain communicate through discrete electrical impulses.
The temporal aspects of spikes not only brings SNNs closer to biological reality but also enables them to process information in a more event-driven, energy-efficient manner \cite{Gerstner2014}.
Despite these advantages, SNNs are not yet widely leveraged for computing applications, among others due to challenges in their training methods \cite{BOHTE2002} and a lack of large scale as well as parallel hardware implementation \cite{TAVANAEI2019,Roy2019}.

Unlike traditional ANN with their instantaneous response according to a nonlinear map, SNNs consist of dynamical units, where neurons' activations evolve over time.
This allows SNNs to process and represent time-varying information inherently, making them especially suited for tasks involving dynamical inputs, such as speech, video, and sensory data.
In SNNs, different encoding schemes can be used to represent information, with two common approaches being rate-based and temporal encoding.
Rate encoding conveys information through the frequency of spikes over a time window, while temporal encoding uses the precise timing of individual spikes.
Temporal encoding schemes offer several advantages: (i) they are more expressive, often requiring fewer spikes to represent the same information, (ii) they are more energy-efficient and faster \cite{VanRullen2001} (iii) they have a natural connection to sparsity \cite{VanRullen2001}. 

At the single-neuron level, numerous neuromorphic devices have been developed to emulate spiking behaviors.
For instance, memristors and spintronic devices have been explored for mimicking neuronal excitability, offering compact, low-power alternatives for spiking units \cite{Wang2018}.
At the larger scale, dedicated hardware like IBM's TrueNorth \cite{merolla2014} and Intel's Loihi chip \cite{davies2018} have demonstrated the potential for CMOS-integrated SNNs, although these systems still face limitations in scalability and parallelism.
A particularly exciting avenue for SNN hardware lies in the field of photonics. Optical systems inherently offer advantages in terms of speed \cite{brunner2013} and parallelism \cite{farhat1985}, as light can propagate and hence can transduce signals with very little latency and energy dissipation \cite{Miller2017}.
In this context, photonic excitable units have been proposed as candidates for mimicking spiking neurons.
Examples include optoelectronic components \cite{Hejda2022}, semiconductor lasers with saturable absorbers \cite{Barbay2011,Selmi2014} or in other compounded configurations \cite{Prucnal2016} as well as optically injected semiconductor lasers \cite{Hurtado2012}.
However, despite these promising developments, no large photonic SNN comprising many individual photonic spiking units has yet been demonstrated, nor an optical SNN that leverages sparsity.

\begin{figure}[t]
	\centering
	\includegraphics[width=1\linewidth]{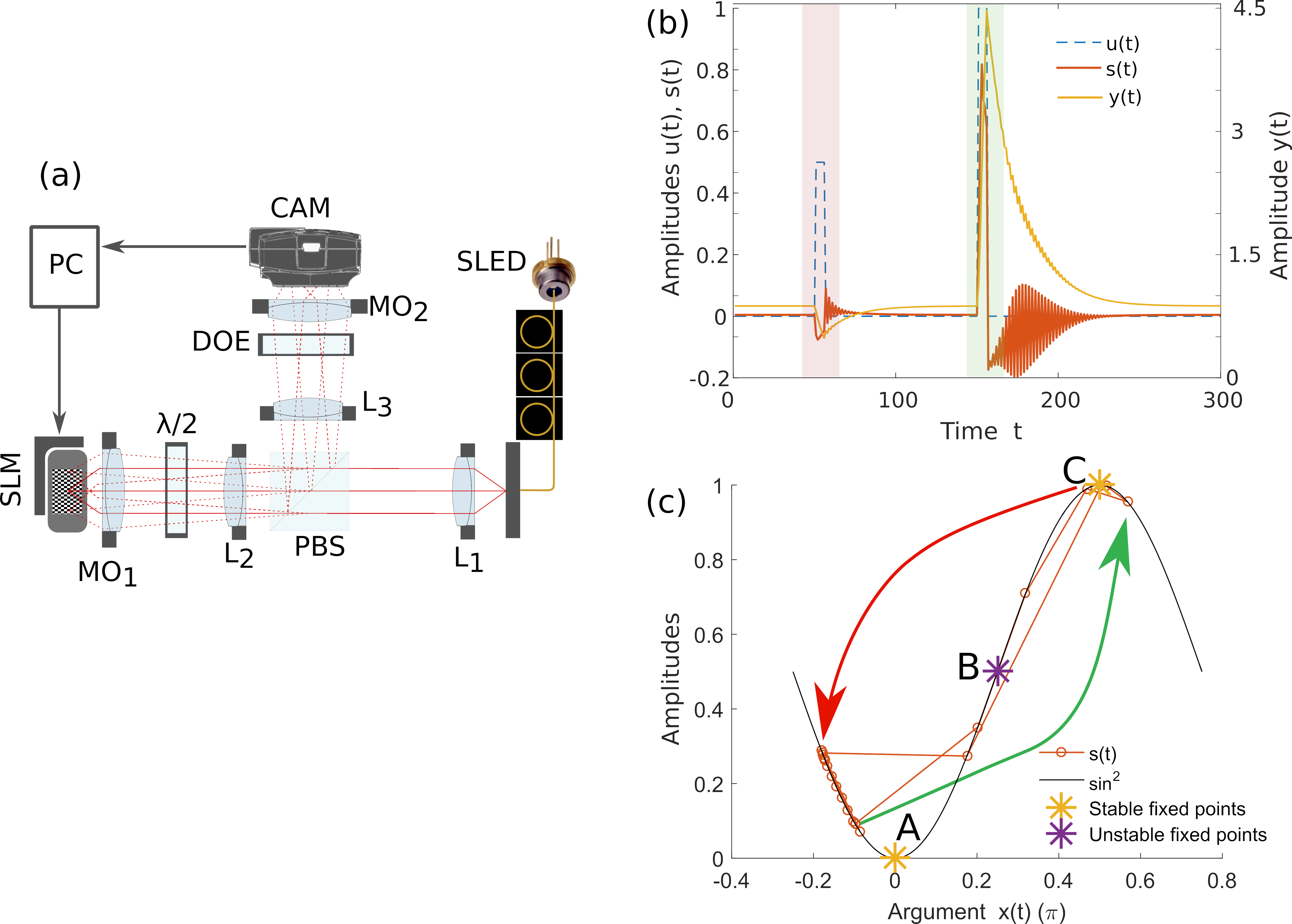}  % Adjust width as needed
	\caption{(a) Schematic of the optical setup.
		The spatial light modulator (SLM) is illuminated by a single-mode fiber coupled superluminescent diode (SLED) that collimated by $\textrm{L}_1$, polarization filters by a polarizing beam splitter (PBS).
		Lens $\textrm{L}_{2}$ and microscope objective $\textrm{MO}_1$ image the SLED's colimated beam, illuminating $\sim40000$ SLM pixels.
		$\textrm{L}_{3}$ and $\textrm{MO}_{2}$ image the SLM on to a camera (CAM), and a diffractive optical element introduced optical coupling between the pixels.
		(b) Excitable dynamics via the high-pass filtered Ikeda map. 
		We apply below excitation strength stimulus via input $u(5:55)=0.5$ (red shaded area) and above excitation strength stimulus $u(150:155)=1$ (green shaded area).
		The neuron's state variable $s(t)$ exhibits an excitable dynamical response only for the second input.
		Slow dynamics $y(t)$ create strong negative forcing only in the second case. 
		(c) Concept of excitability with slow, negative-feedback forcing an Ikeda map.
		At rest, the system resides close to stable fixed-point A.
		If an external perturbation pushes the state past the unstable fixed-point B, the system is attracted to the upper stable fixed-point C, trajectory illustrated by green arrow.
		There, the slow-feedback term builds up until it projects the system back to stable fixed point A, trajectory illustrated by red arrow.
		Parameters were $\gamma=0.3$, $\beta=0.45$, $\delta=0.1$, $\Theta=-0.1\pi$, $\eta=0.995$.}
	\label{fig:ExpSchematic}
\end{figure}

Here, we report the first large-scale photonic SNN comprising of 40.000 neurons, representing a significant step toward realizing a generic proof of concept neuromorphic computing systems based on light.
Our hardware architecture is built using off-the-shelf components, making it both cost-effective and easily scalable to much larger systems.
We employ a liquid-crystal on silicon spatial light modulator (SLM) in combination with a CMOS camera, and both these devices now readily feature more than $10^6$ pixels.
The hybrid nature of the setup, combining digital control with analog photonic processing, offers tremendous flexibility, allowing for straightforward modifications to the governing dynamical equations and adaptability to various computing tasks.
At its core, the system utilizes a photonic response modeled by an Ikeda map \cite{Bueno2018} that corresponds to an iterative map defined by a feedback term and a trigonometric nonlinearity, which provides the necessary nonlinearity and is a prominent model system used to study complex dynamics.
Similarly to \cite{Brunner2018} we introduce a second, slower dimension to the Ikeda map acting as negative and hence inhibitory feedback, which is the first time such a addition to the Ikeda map has been proposed.
This slow-fast, 2D system enables excitability for first time with an Ikeda map, making our photonic neurons respond only when an input exceeds a threshold, akin to biological excitable dynamics.

We conduct the basic characterization of the system and its excitable dynamics, determining its excitability threshold, its excitability type, response latency as well as refractory period.
We then demonstrate its capability to process information through photonic spikes, and we employ two approaches for training it to the MNIST digit recognition task: single-shot learning by ridge-regression for training in the context of liquid state machines (LSM) \cite{Maass2002}, and the hardware-friendly black-box simultaneous perturbation stochastic approximation (SPSA) gradient approximation technique \cite{McCaughan2023}, ensuring efficient optimization without the need for access to internal variables.
It is the first time SPSA is employed to train a hardware neural network, and while here we exclusively train offline, contrary to the LSM-based single-shot training the SPSA training can in principle be directly migrated to photonic hardware.
In a novel approach, we utilize latency encoding, where each neuron computes using only a single spike per computation, maximizing efficiency and minimizing power consumption.
Additionally, we implement sparsity in the network through a Rank Order \cite{VanRullen2001} inspired coding scheme, which sparsifies the SNN activity by only permitting activity for a first neurons to spike, further enhancing the system’s efficiency and speed.
When the system is trained with the SPSA algorithm we find that sparsity aids the SNN's classification accuracy, with the best results of 83.5\% test accuracy achieved using only 22\% of the neurons.
Most importantly, we find that under such conditions we can strongly sparsify the network; using only 8.5\% of all neurons we still achieve a test classification accuracy of 77.5\%.
It is the first time that photonic nonlinearity, excitability, latency encoding and sparsity have been experimentally leveraged for computing, and our proof of concept systems paves the way for future exploration of large-scale photonic neuromorphic systems.

\section{The photonic spiking neural network}
\label{sec:ThePSNN}

Our experiment allows for the optical emulation of a SNN, and the core component is a SLM, which is a dynamically reconfigurable optical device akin to an optical display.
However, instead of emitting light, SLMs are operated with an external illuminating light source, whose amplitude or phase profile they modulate spatially and temporally.
Their spatial modulation happens with discrete pixels of $\sim$10~$\mu$m size, and today SLMs readily host beyond $10^6$ pixels that can be modulated on 50~$\mu$s to 1~s timescales.
This equips SLMs with an astonishing level of parallelism, and as a result, they have driven extensive research in the fields of optical processing and computing \cite{farhat1985,zhou2021}.

The experimental setup is schematically illustrated in Fig. \ref{fig:ExpSchematic}. 
Here, we use a liquid-crystal on silicon SLM (Santec SLM-200) with a pixel pitch of $\sim 8.0\mu$m and a total of $1920\times1200$ pixels.
The illumination source is a super-luminescent diode (SLED, Thorlabs SLD850S-A20W, $\lambda=850~$nm), which is collimated by an achromatic lens ($\textrm{L}_1$, Thorlabs AC254-075-B-ML) and polarization filtered via a polarizing beam splitter (PBS, Thorlabs CCM1-PBS252/M) in whose transmission direction the illumination propagates towards the SLM.
The SLM is operated in the intensity modulation mode by adjusting the illumination's polarization to 45$^{\circ}$ relative to its slow and fast axis via a half-waveplate ($\lambda/2$, Thorlabs AHWP10M-980).
In order to illuminate a large SLM-area, an achromatic lens ($\textrm{L}_2$, Thorlabs AC254-035-B-ML) and a microscope objective ($\textrm{MO}_1$, Olympus LMPLN10XIR) image the SLED's collimated beam onto the SLM surface, creating an illuminated area spanning more than $200\times200$ pixels, and the illumination field for pixel $i$ is $E_i^0$.
While the SLM is generically a 2D plane, we here simplify the notation and use index $i$ to allocate pixels according to their position within the 2D SLM state that is flattened to a vector. 
The optical field reflected off the SLM passes again the PBS to divert the signal to the camera (CAM, IDS UI-3042SE-M), and the associated polarization filtering introduces a nonlinearity according to
\begin{equation}
	\label{eq:SLM}
	E_{i}(t) = E_i^0 \sin\left(2\pi \frac{x_i(t) + \Phi_i}{\kappa_i^{\textrm{SLM}}}\right),
\end{equation}
\noindent where $x_{i}(t)$ is the grayscale value of SLM pixel $i$, $\Phi_i$ is a constant phase offset specific to the device and its pixels; $\kappa^{\textrm{SLM}}$ is a conversion factor relating SLM grayscale to the optical polarization's angle, and $t$ is an integer time.

In order to create coupling between SLM pixels and hence our neurons, a diffractive optical element (DOE) can be positioned between the PBS and the camera.
For more details about this concepts and resulting network topology please see \cite{brunner2015,maktoobi2019}.
Finally, the camera, positioned at the focal plane of the second microscope objective ($\textrm{MO}_2$, Olympus LMPLN10XIR) records the normalized optical intensity
\begin{equation}
	I_{i}(t) \propto | \sum_{j=1}^N W_{i,j}^{\textrm{DOE}} E_{j}(t) |^2,
\end{equation}
\noindent where $\mathbf{W}^{\textrm{DOE}}$ is the optical coupling via the DOE
In practice, the optical intensity is scaled with optical attenuators such that the camera image, normalized between 0 and 1, is maximally leveraging the camera's 8-bit resolution.
Here, the illumination of the SLM's surface is not uniform but follows a Gaussian intensity distribution.
Furthermore, there are small yet notable local differences in phase offset $\mathbf{\Phi}$ and conversion factor $\mathbf{\kappa}^{\textrm{SLM}}$.
In order to account for these non-idealities, we measure the nonlinear transfer function of each pixel individually and determine all 40.000 $I_i^0=|E_{i}|^2$, $\Phi_i$ and $\kappa_i^{\textrm{SLM}}$ through fitting the responses to the squared version of Eq.~\ref{eq:SLM}.
These we then use to normalize the amplitude dynamics, to compensate for the variations in offset and gray-scale to phase change coefficient for each single pixel.

In order to create a nonlinear dynamical map we use the camera state at time $t$ to define the SLM's state at time $t+1$.
This loop is established by a standard digital computer, via which we also add input information $\mathbf{u}(t+1)$ as well as a potential bias term $\mathbf{\Theta}$, creating the governing equations according to
\begin{align}
	\mathbf{x}(t+1) &=  \beta \mathbf{I}(t) + \gamma W^{\textrm{inj}} \mathbf{u}(t+1) + \mathbf{\Theta} , \label{eq:SLMinput}\\ 
	\mathbf{s}(t+1) &= \sin^2(2\pi\frac{\mathbf{x}(t+1)}{\mathbf{\kappa}^{\textrm{SLM}}}). \label{eq:SNNstate}
\end{align}
\noindent The SLM's state is updated at each time $t+1$ using $\mathbf{x}(t+1)$ of Eq.~(\ref{eq:SLMinput}).
The feedback as well as the information input strengths are linearly scaled with  $\beta$ and $\gamma$, respectively, and $\mathbf{W}^{\textrm{inj}}$ are the input connections with randomly distributed values between 0 and 1.
Finally, $\mathbf{\Theta}= \mathbf{\Theta}_0 + \mathbf{\Phi}$ allows including a bias independently for each neuron via the additional phase offset $\mathbf{\Theta}_0$.
These equations correspond to a discrete Ikeda map, which in the past has extensively been used for implementing photonic NNs \cite{Bueno2018}.
As our photonic SNN's state we use $\mathbf{s}(t+1)$ of Eq.~(\ref{eq:SNNstate}) is calculated at each time $t+1$.
In our experiment, the time of one update is 800~ms.
However, in the following we will set time as unit-less since here we report proof of concept results and want to allow for direct comparability between simulations and experiment.

\subsection{Creating excitability}

While the Ikeda map exhibits a wide range of complex behavior, it does not provide excitability.
This is the same for the McCulloch and Pitts neuron in its time-discrete form \cite{LITTLE1974101}, however, by simple addition of an integrating term, such maps can become excitable \cite{GIRARDISCHAPPO,CAIANIELLO}.
In order to achieve this, we amend the internal state of our neurons according to
\begin{align}
	\mathbf{x}(t+1) &=-\delta \mathbf{y}(t)+ \beta \mathbf{I}(t) + \gamma \mathbf{W}^{\textrm{inj}} \mathbf{u}(t+1) + \mathbf{\Theta} , \label{eq:SlowIkeda} \\
	\mathbf{y}(t+1) &= \eta \mathbf{y}(t) + \mathbf{x}(t+1), \label{eq:Memory}
\end{align}
\noindent where $\mathbf{y}(t+1)$ is the second, slow dimension of our system whose retainment of previous states is scaled by $\eta$.
The slow dynamics of Eq.~(\ref{eq:Memory}) act upon our Ikeda system of Eq.~\ref{eq:SlowIkeda}
through negative feedback with strength $\delta$.
Parameters for our following initial evaluation were $\gamma=0.3$, $\beta=0.45$, $\delta=0.1$, $\Theta=-0.1\pi$ and $\eta=0.995$.

\begin{figure}[t]
	\centering
	\includegraphics[width=1\linewidth]{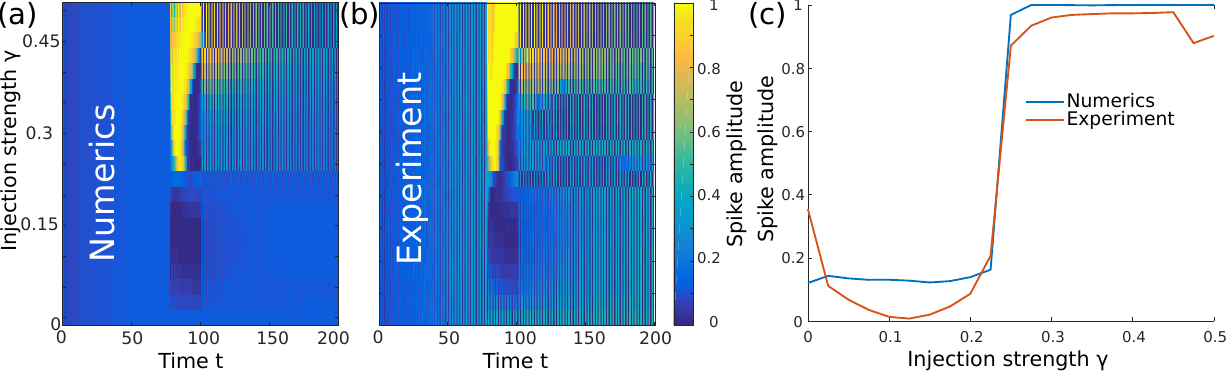}  % Adjust width as needed
	\caption{Response to a short perturbation, with $u(500:end)=1$.
		(a) Excitability of $x(t+1)$ in Eq.~(\ref{eq:SlowIkeda}) as a function of injection strength $\gamma$.
		An excitable response is attained for $\gamma>0.25$.
		(b) Experimental data agrees exceptionally well to the numerical model.
		(c) Maximum amplitude of a response versus $\gamma$ displays the almost ideal all-or-nothing response of the slow-negative feedback forced Ikeda map.}
	\label{fig:Excitable}
\end{figure}

Figure~\ref{fig:ExpSchematic}(b) shows the consequence of slow term $\mathbf{y}(t)$ with negative forcing on the fast dimension $\mathbf{s}(t)$.
First, we perturb the system with a below excitability threshold input $u(50:55)=0.5$ (dashed blue data), and one can see that the response of the system $s(t)<0.1$ remains small.
However, this changes drastically for the following perturbation $u(150:155)=1$, and the system responds with a single, large amplitude spike with $s(t)>0.8$.
The principle underlying our excitability is illustrated in Fig.~\ref{fig:ExpSchematic}(c).
For sufficiently small feedback $\beta$, the system is non chaotic, and we chose offset $\Theta$ such that it resides close and left to its lower stable fixed-point $A$.
From there, external stimulus $\mathbf{u}(t+1)$ can perturb it such that it either remains below or that it passes unstable fixed-point $B$.
If the system remains below $B$, it will directly relax back to around $A$ with some damped oscillatory behavior.
However, should perturbation $u(t)$ push the system passed unstable fixed point $B$, then $s(t)$ will continue towards upper stable fixed-point $C$.
This is where the slow forcing of Eq.~\ref{eq:Memory} starts to matter.
While close to point $A$, the high-pass filtered, i.e. integral term of $y(t)$ remains small.
However, this integral term rapidly grows if the system resides around fixed-point $C$ due to the large $s(t)$ amplitudes.
After a short buildup, $y(t)$ is of sufficient strength to force the system to again passed the unstable fixed-point $B$, from where it relaxes back into its resting state near point $A$.

In Fig.~\ref{fig:Excitable} we compare numerical simulations of Eqs.~(\ref{eq:SlowIkeda},\ref{eq:Memory}) with our experimental findings.
In the experiment we removed the DOE in order to characterize the response of individual neurons that is not perturbed by the impact for SNN topology.
We optimized hyperparameters to $\beta=0.475$, $\Theta=-0.35$, $\delta=0.1$ and $\eta=0.995$, and here we subject the system to a single perturbation of $u(50:75)=1$.
Figure~\ref{fig:Excitable}(a,b) show the response of the system for a range of $\gamma\in [0,0.5]$ for the numerical model and the normalised experiment for one neuron $i=20000$, respectively.
For too small injection strength $\gamma$, the system is not pushed across unstable fixed point $B$, and no excitable response is obtained.
As soon $\gamma$ exceeds this threshold, which for this set of hyperparameters is at $\gamma\approx0.23$, the system does a full amplitude excursion, passing from unstable fixed point $B$ to $C$, followed by forcing the system back again to stable fixed point $A$.
Figure~\ref{fig:Excitable}(c) shows the maximum amplitude to better illustrate the generic all-or-nothing response of an excitable neuron for, both, the numerical model and the normalized experiment for which we averaged all 40.000 SNN responses.
Here, we see the damped oscillations following a perturbation that were observable in Fig.~\ref{fig:ExpSchematic}(b) in detail.
They are more pronounced in the experiment, which we assign to an asymmetry of the SLM's nonlinear function that varies from pixel to pixel.

\subsection*{Excitability type and refractory period}

\begin{figure}[t]
	\centering
	\includegraphics[width=1\linewidth]{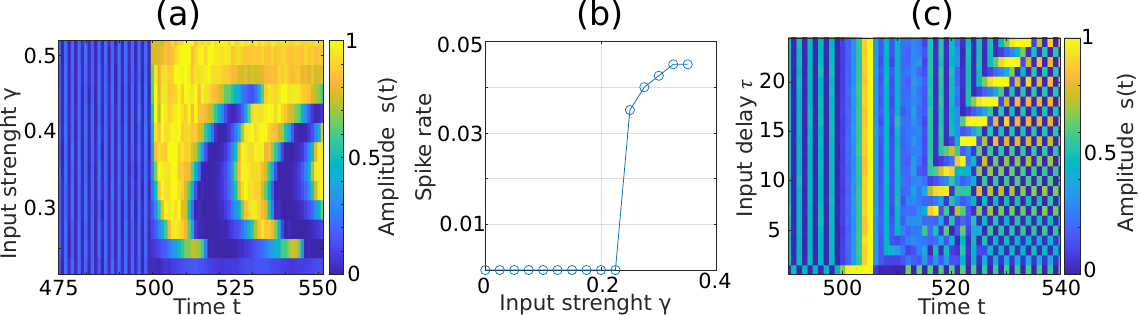}  % Adjust width as needed
	\caption{Spike rate and refractory period.
		(a) Continuous stimulation with $u(500:\textrm{end})=1$ results in a continuous  spike train as soon as the excitation threshold is crossed.
		(b) The spike rate as a function of injection strength $\gamma$ exhibits classical type 1 excitability characteristics.
		(c) Numerical simulation.
		After an initial excitation with $u(500:505)=1$ we subject the neuron to a second spike at time $u(506+\tau : 510+\tau)=1$ at $\gamma=0.3$.
		For $\tau<12$ the neuron cannot be re-excited, demonstrating the refractory period this modified Ikeda map.
		All data are from the experiment.}
	\label{fig:SpikeRate}
\end{figure}

In the following we further inspect the nature of our photonic spiking neurons.
We here exclusively show experimental data, however, each data are exceptionally well reproduced by the numerical model.
We start by subjecting the photonic SNN to a constant input stimulus according to $u(500:end)=1$, and Fig.~\ref{fig:SpikeRate}(a) shows the representative response of a single photonic neuron.
Starting from a threshold around $\gamma=0.27$ the neuron abruptly starts to exhibit spike train responses with a spike-rate of$\sim 0.035$ which increases with $\gamma$ until it saturates at 0.45, see Fig.~\ref{fig:SpikeRate}(b).

The excitability behavior of neurons is broadly categorized by two classes of responses.
Neuron excitability type 1 refers to a response where a neuron can fire spikes at arbitrarily low frequencies when the excitability threshold is exceeded.
It hence exhibits a continuous transition from no spiking to spiking, meaning the spike rate increases smoothly from zero as the stimulus intensity increases.
It is commonly associated with neurons that exhibit a saddle-node bifurcation mechanism in their firing dynamics.
Neurons with excitability type 2 exhibit a sudden jump to a non-zero firing rate when the excitability threshold is exceeded.
In contrast to Type 1, neurons hence start spiking at a finite frequency.
This behavior is associated with a Hopf bifurcation in the neuron’s dynamics.
According to the experimental and numerical data of our photonic SNN exhibits dynamics according to Type 2 excitability.

Another characteristic property of neurons are their refractory period.
The refractory period refers to the time following an preceding spike during which a neuron is less or not excitable.
Here, one again has to differentiate between two distinct behaviors.
An absolute refractory period corresponds to a time during which a neuron is not excitable, regardless of the strength of an input stimuli.
During a relative refractory period the neuron can fire again, but only if the incoming stimulus is stronger than usual.
To investigate our photonic SNN's refractory behavior we subject its neurons to an initial input with $u(500:505)=1$, which is followed by a second stimulus that is delayed by $\tau$ according to $u(506+\tau:510+\tau)=1$.
In Fig.~\ref{fig:SpikeRate}(c) we show the response of a neuron as we tune input delay $\tau$, and this response is representative of all neurons in the network.
We can clearly see that after the initial spike the neuron is incapable of spiking again for a delay $\tau<8$.
Importantly, when the strength of the second stimulus is raised to $u(506+\tau:510+\tau)=2$ then this refractory period window disappears entirely.
Our photonic SNN therefore harbors neurons that exhibit a relative refractory period of 7 time steps, yet they do not exhibit an absolute refractory period.

\subsection*{Spike latency}

\begin{figure}[t]
	\centering
	\includegraphics[width=1\linewidth]{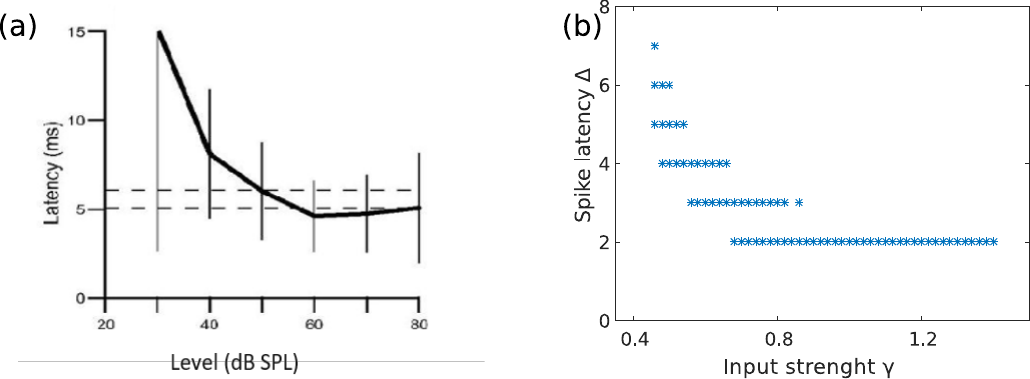}  % Adjust width as needed
	\caption{between applying an above threshold perturbation and a neuron's response.
		(a) Typical spike latency of a biological neuron, data from \cite{recio2014}.
		(b) Our experimental results of spike latency $\delta$ exhibits very similar behavior.}
	\label{fig:Spikelatency}
\end{figure}

As the final fundamental characterization we analyze the latency in our photonic neurons' responses as a function of input strength $\gamma$.
In biological neurons a stronger stimulus usually results in a shorter latency between the input and the neuron's spiking response, while weaker stimuli may lead to longer delays.
Figure~\ref{fig:Spikelatency}(a) shows an example for cat neurons in an auditory structure called the central nucleus of the lateral lemniscus, for which the spike-latency drops from 15 to 5~ms over the dynamic range of the neuron's input.
Such spike latency is an important mechanism for coding concepts that use temporal neuronal coding strategies with a particular relevance for SNN concepts that leverage sparsity through rank-order coding concepts.
For our photonic SNN we characterized latency $\Delta$ between spike-response and input stimulus as a function of injection strength $\gamma$, and our hyper parameters were same as mentioned previously.
As a neuron's time we heuristically defined the moment when its amplitude crosses threshold $s(t)>0.6$, and the photonic SNN's latency response curve $\Delta(\gamma)$ for all 40.000 neurons is shown in Fig.~\ref{fig:Spikelatency}(b).
Just as for the biological neuron, $\Delta(\gamma)$ follows an exponential decay, here dropping from $\Delta(0.42)=7$ to $\Delta(\gamma > 1)=2$, which in turn enables the photonic SNN to latency-encode its response to an injected information with a resolution equivalent to 2.8 bits.

\section{Sparsity via rank order coding for MNIST digit recognition}
\label{sec:SparsityMNIST}

\subsection{The sparse photonic SNN}

A clear discrepancy between biological brains and current NN concepts is that the brain leverages extreme sparsity, with only a fraction of neurons spiking for each reaction to sensory stimulus \cite{vinje2000sparse}.
While the precise algorithm leveraged by brains remains still unclear, one path to exploit sparsity in neuron activity is using spike latency \cite{VanRullen2001}. 
Using latency as information encoding state variable, neurons' spike responses in a SNN are ranked according to their spike latency $\Delta$.
As shown in Fig.~\ref{fig:Spikelatency}(b), latency is directly related to the strength of a neuron's input, and as such rank order coding via $\Delta$ comes with highly attractive prospects.
Higher ranked neurons correspond to the ones receiving the strongest input, establishing a straight forward mechanism to map an input's hierarchy to a time window in a SNN's response.
This allows for a direct avenue for information compression, as shown for the example of image reconstruction \cite{van2001rate}.
This temporal hierarchy opens the possibility to easily enforcing sparsity by only allowing a certain number of latencies and hence neurons to spike before using lateral inhibition to quench the SNN's activity as a whole.

We apply the same concept to the MNIST handwritten digits as input information.
The so far scalar input $u(t)$ is now replaced with a vector containing the flattened MNIST images, and as stated in Eq.~\ref{eq:SlowIkeda}, input connectivity matrix $\mathbf{W}^{\textrm{inj}}$ was randomly initialized between -1 and 1 and normalized by its largest eigenvalue.
For 23 time steps, input $\mathbf{u}(t)$ is associated to the same image in order to give the photonic SNN time to respond.
Furthermore, sequential inputs are separated by 25 time steps during which the input is clamped to zero, which prevents the content of sequential images perturbing the response due to transient dynamics in slow variable $\mathbf{y}(t)$.
However, this directly illustrates the potential of our concepts to process the context of sequential images, i.e. movies.

\begin{figure}[t]
	\centering
	\includegraphics[width=1\linewidth]{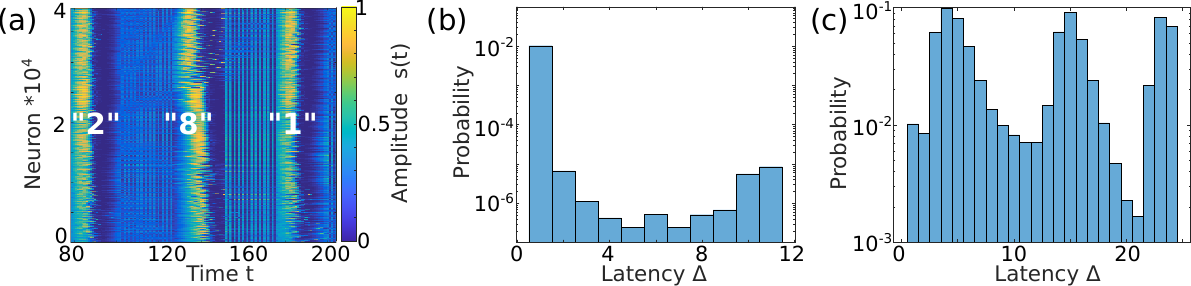}  % Adjust width as needed
	\caption{(a) Spatio-temporal spike pattern responses of our photonic SNN's to MNIST examples of digits 2, 8 and 1.
		Histogram of spike latencies $\Delta$ for (b) $\Delta^{\textrm{l}}=1$ and (c) for $\Delta^{\textrm{l}}=23$.
		This corresponds to only allowing the earliest or neurons to spike, respectively, resulting in corresponding sparsities of 98.98\% and 18.5\%.}
	\label{fig:MNISTlatency}
\end{figure}

Figure \ref{fig:MNISTlatency}(a) shows temporal response of all neurons in our photonic SNN to the injection of three different MNIST examples of digits 2, 8 and 1.
This spatio-temporal response clearly reveals how latency encoding translates the purely spatial features of the different MNIST images into temporal features imprinted on a diverse spatial distribution of latency delays $\mathbf{\Delta}$.
The sparsity-inducing procedure is the following.
We start counter $c$ as soon as the first neuron spikes in response to an input at time $t_0$, and $c$ increases until it reaches its limit $\Delta^{\textrm{l}}$ after which we switch the photonic SNN's energy supply off:
\begin{align}
	c(t) &= t-t_0, \label{eq:Counter} \\
	\mathbf{E}^0|_{c(t)>\Delta^{\textrm{l}}} &= 0. \label{eq:Quenching}
\end{align}
\noindent We hence allow the photonic SNN to evolve according to Eqs.~(\ref{eq:SlowIkeda},\ref{eq:Memory}) until $c(t)>\Delta^{\textrm{l}}$.
At time $t>t_0+\Delta^{\textrm{l}}$ we would switch the illuminating SLED diode off, corresponding to strong lateral inhibition of all photonic SNN neurons.
This takes the equivalent role of clamping the photonic SNN off its energy supply, and hence the energy consumption of the SNN would be significantly reduced.

However, in our proof of concept experiments we let our SNN run freely to record all spiking responses, and we emulate the effect of such sparsity through lateral inhibition simply by only considering neurons that have spiked a times $c(t) \leq \Delta^{\textrm{l}}$; all other neurons are ignored for the following analysis.
Figure \ref{fig:MNISTlatency}(b) and (c) show the photonic SNN's spike latency distribution histograms for setting $\Delta^{\textrm{l}}=1$ and $\Delta^{\textrm{l}}=23$, respectively.
For these values, 1.02\% and 81.5\% of neurons are allowed to spike, corresponding to 98.98\% and 18.5\% sparsity.
The significant diversity of temporal responses in our photonic SNN can be appreciated from the wide distribution of spike delays in Fig.~\ref{fig:MNISTlatency}(b).
There, we only allow the narrowest latency window of $\Delta^{\textrm{l}}=1$, yet measured across 6060 MNIST examples, latencies are spread across 12 time steps.

\subsection{Using the sparse photonic SNN for MNIST digit classification}

In our final benchmark evaluation, we train our photonic SNN to classify the hand written digits of the MNIST benchmark test in the context of a LSM, i.e. only training the photonic SNN's output weights $\mathbf{W}^{\textrm{out}}$ in an offline procedure.
A sparse network response was used as the input, with sparsity defined by the sequential inclusion of neurons ranked by their spike delay.
The input dataset was initially formed by selecting neurons with the shortest spike delays. Neurons with the next shortest delays were then incrementally added, followed by those with increasingly longer delays.
This process was repeated, resulting in a series of sparse network responses, each of which served as an input.
As state-vector for creating our photonic SNN's prediction we used each neuron's amplitude once it crossed $s_{i}(t)>0.6$, which we defined as threshold for registering a neuron's response as a spike.
As loss function $L$ we use the normalized mean square error, and as training routing we restrict ourselves to a hardware-friendly gradient estimation technique called Simultaneous Perturbation Stochastic Approximation (SPSA).
This stochastic gradient descent based method was introduced by J.C. Spall in 1987 \cite{spall1987}, and it is excellently suited for hardware-based optimization as per epoch it only requires two performance evaluations to estimate the gradients for all weights.
It is thus a computationally highly efficient algorithm that has been used to optimize simple NN controllers \cite{choy2004}.
Each training epoch, the SPSA algorithm perturbs and updates $\mathbf{W}^{\textrm{out}}$ according to
\begin{align}
	\mathbf{g} &= \frac{L(\mathbf{W}^{\textrm{out}}+\epsilon\Lambda) - L(\mathbf{W}^{\textrm{out}}-\epsilon\Lambda)}{2\epsilon\textrm{VAR}(\Lambda)}\Lambda\approx \nabla L(\mathbf{W}^{\textrm{out}}), \label{eq:SPSAgradient} \\
	\mathbf{W}^{\textrm{out}} &\leftarrow \mathbf{W}^{\textrm{out}} - \eta \mathbf{g}. \label{eq:Quenching}
\end{align}
\noindent Here, $\epsilon=1/2^{10}$ is a small constant that here we set to the resolution of a 10-bit SLM.
Furthermore, $\Lambda$ is a vector comprising elements randomly drawn at each training epoch from the two integers +1 and -1.
After hyperparameter optimization we set the learning rate to $\eta=10^{-4}$.

\begin{figure}[t]
	\centering
	\includegraphics[width=1\linewidth]{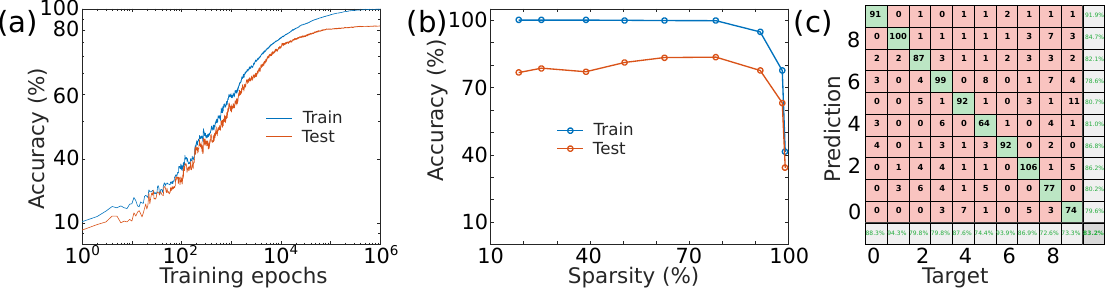}  % Adjust width as needed
	\caption{(a) Testing accuracy convergence during SPSA training, using a sparsity of 78.06\%.
		(b) We achieve a constant training accuracy of 100\% until a sparsity of 78.06\%.
		Most importantly, the testing accuracy is highest at the same high sparsity, and even a network operated with 98.08\% sparsity still achieves 63.11\% testing accuracy.
		(c) Confusion matrix for the highest test accuracy achieved with 78.06\% sparsity.}
	\label{fig:MNISTtask}
\end{figure}

We trained our system using 5000 random examples of the MNIST data set, while we used further 1060 examples for testing that have not been part of the training dataset.
Figure~\ref{fig:MNISTtask}(a) shows the convergence of our photonic SNN's training and testing error when using an inhibition at $\Delta^{\textrm{l}}=3$, at which only 21.94\% of all neurons spike, corresponding to a sparsity of 78.06\%.
From training our photonic SNN we find that our system excellently learns the data of the training set, reaching 100\% training accuracy after $\approx10^5$ training epochs.
However, the testing performance is slightly lower, reaching 83.49\%.
The remains below the limit of a linear classifier ($\approx93\%$), and this slightly low testing performance we assign firstly to the still limited size of our training data set, and secondly to the limit resolution in spike latency of only 2.8 bit.

Our most important finding is however the excellent performance of our photonic SNN in the context of sparsity, see experimental data in Fig.~\ref{fig:MNISTtask}(b).
The first relevant finding is that sparsity helps the SNN's expressivity, and our testing accuracy systematically increases from 76.6\% to 83.49\% when we augment the photonic SNN's sparsity from 18.5\% to 78.06\%, respectively.
The second and most astonishing outcome is that testing accuracy remains very robust against even more radical enforcing of sparsity.
For sparsities of 91.42\% and even 98.08\% we still reach testing accuracies of 77.55\% and 63.11\%!
This opens powerful approaches to strongly reducing a network's energy consumption by adjusting the allowed latency window $\Delta^{\textrm{l}}$ to the difficulty of the task.
Finally, Fig.~\ref{fig:MNISTtask}(c) shows the confusion matrix for our best operation conditions at 78.06\% sparsity.

\section{Conclusion}
\label{sec:Conclusion}

In our work we have experimentally demonstrated the implementation of an opto-electronic SNN comprising a record of 40.000 neurons.
Our proof-of-concept photonic SNN is exclusively based on off-the-shelf components, making it a prime candidate for wide exploration in photonic research.
Using a LCOS-SLM operated in intensity modulation we make the system excitable by introducing a slow variable via the digital control computer before closing the loop and sending the state recorded by a CMOS camera back to the SLM.
Based on this experiment we are able to identify and characterize several key metrics of biological neurons, such as excitability type (here type 2), a relative refractory period of 6 times steps as well as an exponentially decaying spike latency.

These attractive features we then leverage to study SNN computing accuracy in the context of network sparsity.
We systematically enforce sparsity in our photonic SNN's response by leveraging a lateral inhibition mechanisms that is gated through time window $\Delta^{\textrm{l}}$ in the context of spike-latency driven rank order coding.
Tuning $\Delta^{\textrm{l}}$ from 1 to 23 dramatically changes the photonic SNN's sparsity when injected with MNIST hand written digit data.
For the shortest sparsity window with $\Delta^{\textrm{l}}=1$ only 1.02\% of all neurons spike, corresponding to a sparsity of 98.98\%, while for $\Delta^{\textrm{l}}=23$ these numbers correspondingly increase to 81.5\% spiking neurons and 18.5\% sparsity.
Most fascinating in our finding is however the effect sparsity has on our photonic SNN's classification accuracy.
Firstly, we find that sparsity has a beneficial impact upon the system's accuracy, which is expected due to associated increase in response expressivity, and the best classification test accuracy of 83.49\% is achieved at 78.06\% sparsity.
Secondly, via the simple gating mechanism introduced through $\Delta^{\textrm{l}}$ one now has a direct mechanism at hand that enables to almost continuously optimize between the trade-off of task-accuracy versus response time and energy consumption.
The fastest and most efficient reply, i.e. for $\Delta^{\textrm{l}}$, only requires counting the very first neurons to spike, corresponding to 1.02\% network activity or 98.98\% sparsity.
With this fast and highly efficient network the photonic SNN still achieves a testing accuracy of 34.2\%, more than three times above chance.
This is certainly far away from perfect, yet it allows ultra-fast responses and efficiency, e.g. in the context reflex-like fight-or-flight responses for artificial systems.
Simply by increasing $\Delta^{\textrm{l}}$ one increases the system's accuracy in conditions where time and energy are not limited resources.
This is the first time such a powerful attention and resource regulating mechanism has been demonstrated in a photonic SNN.

Finally, we would like to point out that the large size of our photonic SNN is fundamentally required for efficient rank-order coding.
This temporal information embedding concept fundamentally relies on sparsity, making use of the increasing expressivity of geometric patterns when only a small subset of states is allowed to be active.
Simply replacing the 40.000 neuron's we have implemented with the 8560, corresponding to the activity of the best performing 78.06\% sparsity, would hence not create the same performance.

\begin{acknowledgments}
	
	This work was supported by the Agence Nationale de la Recherche (ANR-21-CE24-0018-02); European Research Council (Consolidator Grant INSPIRE, 101044777); European Union Horizon research and innovation program under the Marie
	Sklodowska-Curie Doctoral Training Networks (860830, POST DIGITAL) .

\end{acknowledgments}

\section*{References}
\bibliography{bibliography}% Produces the bibliography via BibTeX.
	
\end{document}